\documentclass[sigconf]{acmart}

\AtBeginDocument{%
  }

\setcopyright{acmlicensed}
\copyrightyear{2018}
\acmYear{2018}
\acmDOI{XXXXXXX.XXXXXXX}
\acmConference[Conference acronym 'XX]{Make sure to enter the correct
  conference title from your rights confirmation email}{June 03--05,
  2018}{Woodstock, NY}
\acmISBN{978-1-4503-XXXX-X/2018/06}

\usepackage{multirow}
\usepackage{subfig}
\usepackage[font=normalsize,labelfont={bf}, skip=1pt]{caption}
\usepackage{booktabs}
\usepackage[table]{xcolor}
\usepackage{threeparttable} 
\usepackage{booktabs}
\usepackage[table]{xcolor}
\usepackage{threeparttable}
\usepackage{tabularx}
\usepackage{ragged2e} 
\usepackage{enumitem}

\begin{document}

\newcommand{\rqone}{How does suppressing phone notifications affect the number of breaks CS1 students take while programming?}
\newcommand{\rqtwo}{Does the effect of suppressing phone notifications vary by assignment?}

\title{Student programming behavior with and without phone notification suppression}





\author{Gavin Eddington}
\affiliation{%
  \institution{Utah State University}
  \city{Logan}
  \state{Utah}
  \country{USA}
}

\author{Christopher Warren}
\affiliation{%
  \institution{Utah State University}
  \city{Logan}
  \state{Utah}
  \country{USA}
}

\author{Seth Poulsen}
\affiliation{%
  \institution{Utah State University}
  \city{Logan}
  \state{Utah}
  \country{USA}
}

\author{John Edwards}
\affiliation{%
  \institution{Utah State University}
  \city{Logan}
  \state{Utah}
  \country{USA}
}

\renewcommand{\shortauthors}{Trovato et al.}






\begin{abstract}

\textbf{Background and Context}. Computer programming often involves extended periods of sustained activity and mobile phone notifications introduce frequent opportunities for interruption. Prior work demonstrated that keystroke-derived breaks can serve as an indicator for disengagement and provided preliminary evidence that suppressing phone notifications may reduce these disruptions.

\textbf{Objectives}. Our primary research question is: How does suppressing phone notifications affect students’ task engagement and productivity while programming?

\textbf{Method}. We report on a replication and methodological extension study conducted in a CS1 course involving 22 students. Using a within-subject design, selected programming assignments were randomly designated for enabling notification suppression. Phone state logs (Android and iOS) were synchronized with millisecond-resolution IDE keystroke data to measure student attention and focus when in the control and notification-suppression conditions.

\textbf{Findings}. Assignments completed with notification suppression enabled significantly lower break rates and longer intervals of focus compared to assignments completed in the control condition for many, but not all, students. Mixed-effect modeling revealed a general population-level reduction in observed breaks under notification suppression, although the magnitude of effect varied across students. This study provides evidence that notification suppression is associated with measurable differences in programming engagement and behavior. We also find a remarkable bimodality in the effect across students -- many students are positively affected, a small number are negatively affected, and very few experience little or no effect. This finding is consistent with other studies in diverse disciplines.

\textbf{Implications}. Our results show that, for many students, phone notification suppression tools, such as Do Not Disturb, can improve attention and focus. This is a straightforward, low-effort, and scalable intervention that can be suggested by instructors and incorporated by students who are attentive to their own self-regulatory practices. Implications extend to scholarship, as our understanding and measurements of the effects of phone distraction continues to evolve.

\end{abstract}



\maketitle

\section{Introduction}
Keystroke data have enabled researchers to examine programming engagement in CS1 contexts. Prior work has shown that prolonged pauses in typing behavior can serve as indicators of interruption or disengagement during programming tasks ~\cite{hart2023accurate}. These methods provide insight into when students pause, but they do not directly capture the potential sources of interruption, such as mobile device notifications.

Students frequently complete programming assignments while surrounded by digital distractions. Text messages, social media alerts, and other application notifications introduce disparate demands for their attention ~\cite{stothart2015attentional}. Even when not explicitly acted upon, notifications can impose cognitive costs that disrupt problem solving ~\cite{pielot2017productive, ward2017brain, thornton2014mere}. Despite widespread concern about phone-level distractions~\cite{martin2025digital, chen2025mobile, deng2025smartstudents, Saad_2023}, little  work has connected phone states to observable programming behavior while completing homework assignments. Even less work has been done when observations are taken in a natural setting rather than a laboratory setting.

A prior study examined six CS1 students who alternated between using Do Not Disturb (DND) and not using it while completing programming assignments ~\cite{hart2024phone}. Using keystroke logs, that work found preliminary evidence that DND was associated with reductions in long pauses and estimated a productivity gain of approximately $5-12\%$ for several participants. However, the study was limited by a small number of engaged participants, large variability in adherence to using DND, and largely subject-level statistical analyses. The prior work provided suggestive evidence that suppressing notifications may reduce disengagement, but it could not determine if the effect generalizes across a broader population of students or if it varies extensively by task.

This study builds on that foundation by scaling the design to a substantially larger cohort, including 22 students with usable system-level phone data and 47 additional control students with keystroke logs. We extend the original methodology by adding iOS support, implementing stronger notification suppression tracking via device-level controls, and analyzing a broader set of assignments. This larger dataset allows us to move away from individual subject comparisons and estimate population-level effects that account for student and assignment variability.



Our research questions for this study are:
\begin{itemize}
\item[\textbf{RQ1}] \rqone{}
\item[\textbf{RQ2}] \rqtwo{}
\end{itemize}

\section{Related Work}
\label{sec:relatedWork}

\subsection{Task Engagement and Time-on-Task}

Time-on-task (TOT) has long been identified as a critical factor in learning. Foundational work argues that learning outcomes are strongly related to the amount of time learners spend actively engaged with instructional tasks~\cite{carroll1963model, bloom1974time, stallings1980allocated, karweit1984time}. Chickering and Gamson summarized this relationship as ``time plus energy equals learning''~\cite{chickering1989seven}. Importantly, time-on-task reflects not merely elapsed time, but time spent meaningfully engaged, with cognitive effort directed toward the task at hand.

Computer science is a particularly well-suited domain for studying engagement and time-on-task because programming activities allow tracing fine-grained interactions. Prior work has shown that keystroke data can be used to model student engagement during programming tasks and to estimate time-on-task with high accuracy in authentic classroom settings ~\cite{edwards2022practical, hart2023accurate, hart2024phone}. Beyond engagement estimation, keystroke data has been used to study plagiarism~\cite{hellas2017plagiarism, hart2023plagiarism}, identify programmers from typing patterns~\cite{longi2015identification}, and characterize differences in programming behavior across contexts~\cite{edwards2020study}. The resolution of keystroke data enables observation of short pauses and longer disengagements that are difficult to capture through self-report or coarse time measures.

\subsection{Keystroke-Based Measures of Programming Engagement}

Recent work has formalized keystroke-derived measures of engagement by modeling pauses in programming activity. Edwards et al.~\cite{edwards2022practical} introduced a practical model of student engagement based on programming breaks, demonstrating that break behavior is meaningfully associated with student outcomes. Subsequent studies refined these measures and validated keystroke-based engagement metrics across courses and institutions~\cite{hart2023accurate, leinonen2022time}. Related work has also examined pausing behavior as a lens for understanding cognitive load and task difficulty in CS1 assignments~\cite{urry2024framework}.

The present study builds directly on this line of work, using similar IDE instrumentation and break-based engagement measures. Whereas prior studies focused on modeling engagement from keystroke data alone, this work extends existing methods by incorporating system-level phone state information. By aligning keystroke-observed programming behavior with periods in which phone notifications were enabled or suppressed, this study examines whether notification management strategies are associated with observable differences in programming engagement.

\subsection{Phone Notifications and Attention}

A substantial body of research characterizes phone notifications as sources of attentional disruption. Experimental studies have shown that receiving or even anticipating notifications can impair performance and increase task-irrelevant thought~\cite{end2009costly, stothart2015attentional}. Notifications have been associated with reduced learning in classroom settings~\cite{graben2022receiving} and decreased productivity in work contexts~\cite{kim2016timeaware, pielot2017productive}. Importantly, notifications can be disruptive even when they are not acted upon, as they may capture attention or prompt internal monitoring for missed information~\cite{stothart2015attentional, fortin2022understanding}.

More recently, a meta-analysis synthesizing randomized and experimental studies found that mobile phone distraction has a statistically significant negative effect on immediate learning recall among young adults~\cite{chen2025mobile}. By aggregating evidence across controlled studies, this work provides quantitative support for the claim that phone-based multitasking impairs cognitive performance in academic contexts.

Cognitive psychology research suggests that such effects arise from task switching and attentional capture. Switching attention away from a task incurs a cognitive cost, as working memory resources must be reallocated and task context reconstructed~\cite{rogers1995costs}. Stimuli that are personally salient, such as phone notifications, are particularly likely to capture attention involuntarily~\cite{moray1959attention, roye2007personal}. While these effects have been demonstrated across a variety of laboratory and applied settings, less work has examined how notification-related distraction manifests in fine-grained behavioral data during extended, self-paced programming tasks.

Recent large-scale randomized controlled trials further demonstrate that the effects of smartphones in classrooms depend critically on how they are integrated into instruction. Deng et al.~\cite{deng2025smartstudents} conducted two RCTs comparing smartphone bans, unrestricted use, and teacher-guided instructional use. Allowing smartphones without guidance reduced performance relative to bans, whereas guided instructional use significantly improved performance gains. Decomposing time spent learning versus being distracted showed that even modest productive smartphone use can outweigh distraction costs. However, this work focuses on lecture-based verbal instruction and less is known about how notification management affects engagement during extended, self-paced programming tasks.

\subsection{Do Not Disturb and Notification Management}

Do Not Disturb (DND) is a system-level feature designed to suppress non-essential notifications. Prior research on DND has primarily examined its use over extended periods. Pielot et al.~\cite{pielot2017productive} found that participants reported feeling both more productive and more anxious during a 24-hour no-notification period, with individual differences in perceived benefit and stress. Other work has shown that muting notifications does not always reduce phone use and may even increase compensatory checking behaviors, particularly among users concerned about missing important information~\cite{liao2022sound, yoon2015study, fitz2019batching}.

Research examining DND usage at the level of individual tasks is comparatively limited. Zamani et al.~\cite{zamani2019effectiveness} studied DND strategies in a clinical documentation context and found reductions in errors, suggesting potential benefits in interruption-sensitive tasks. 


\subsection{Phone Addiction and Habitual Checking}

The pervasive integration of smartphones into daily life has led to increased prominence of phone-related stimuli. Notifications and the expectation of communication can prompt habitual checking behaviors and attentional monitoring, even in the absence of explicit alerts~\cite{ward2017brain, kruger2016high, kruger2017bad}. Feelings associated with fear of missing out (FOMO) and social obligation have been linked to increased distraction and phone use~\cite{przybylski2013motivational, oberst2017negative}. These factors may complicate the effects of notification suppression, as users differ in their responses to reduced connectivity.

Together, prior work suggests that while notification management tools such as DND have the potential to reduce interruption, their effects are context-dependent and shaped by individual differences. The present study contributes to this literature by examining how DND usage during programming assignments relates to fine-grained behavioral measures of engagement, rather than relying on self-reported productivity or outcome measures.

\subsection{Digital Well being Tools, Restriction Design, and Compliance}

Recent systematic work has examined digital distraction more broadly in educational settings, synthesizing causes, consequences, and intervention strategies across devices and modalities~\cite{martin2025digital}. This review highlights the multifaceted nature of digital distraction, emphasizing that technological affordances, individual differences, and contextual factors jointly shape how devices influence engagement and learning outcomes.

Beyond simple notification silencing, modern mobile devices provide digital well being tools that allow users to restrict access to specific applications or categories of apps. Examples include Apple's Screen Time app limits and downtime features, as well as Android's Digital Well being and Focus Mode systems. These tools function not only by suppressing notifications, but by introducing friction or temporary access restrictions designed to reduce habitual checking behavior.

Research suggests that such restriction-based tools can influence smartphone use, though effects are heterogeneous and often shaped by user motivation and compliance. Hoong~\cite{hoong2021selfcontrol} examines the adoption of iOS Screen Time app limits and finds evidence that usage restrictions can reduce time spent in targeted applications. A broader systematic review of applications designed to reduce mobile phone use identifies built-in operating system tools, including Screen Time, among interventions that have demonstrated reductions in phone use under certain conditions~\cite{rahmillah2023apps}. 

At the same time, prior work highlights the importance of compliance and circumvention. Users may override, disable, or compensate for restrictions depending on perceived importance of notifications or fear of missing out~\cite{liao2022sound, pielot2017productive}. Research on digital interruptions similarly emphasizes that suppression strategies may alter patterns of checking behavior rather than eliminate them entirely~\cite{ward2017brain}. This suggests that the effectiveness of interruption management tools depends not only on their technical design, but also on how users interact with and adhere to them in practice.

The present study aligns with this line of work by examining the use of system-level restriction mechanisms during programming assignments. Rather than enforcing mandatory blocking, students retained control over whether restrictions were enabled, allowing investigation of interruption management under realistic compliance conditions in an educational setting.


\section{Methods}
\label{sec:methods}

\subsection{Context and Participants}

Data were collected during the Fall 2025 semester in an introductory programming (CS1) course at a midsize U.S. university. Our study builds on prior keystroke-based programming research, using similar IDE instrumentation with our own mobile app for phone logging and notification suppression.

Participants installed a custom mobile application that blocks phone notifications when in the foreground. Additionally, it logs events to a server when it enters the foreground or background and when the phone is locked or unlocked. Participants phones are considered to be in notification suppression mode when the app is in the foreground or when the phone is locked.


\subsection{Data Collection}
Programming activity was recorded on students’ laptops using the ShowYourWork (SYW) PyCharm IDE plugin, which has been used in prior work~\cite{showyourwork, hart2024phone}. SYW runs locally within the PyCharm development environment and logs fine-grained editor events such as keystrokes, edits, and paste actions. All events were timestamped with millisecond resolution.

Phone notification suppression usage was recorded on students’ mobile devices using our custom application, which supports both Android and iOS platforms. This extends earlier work that relied solely on Android instrumentation~\cite{hart2024phone}. The mobile application logs timestamps corresponding to the activation and deactivation of system-level notification suppression modes. No information about phone applications, notifications, or message content was collected.

Students in the experimental group installed both the SYW PyCharm plugin and the mobile suppression application. Students in the control group installed only the SYW plugin and did not install the mobile application. As a result, keystroke data were available for both groups, while phone suppression state data were available only for experimental participants.

Keystroke and phone data streams were aligned using timestamps to correlate programming activity with notification suppression status.




\subsection{Study Design}

We used a within-subject design, for each assignment. Across 8 assignments, study participants were asked to use the notification suppression app (the experimental condition) for approximately half, and not enable suppression (the control condition) for the remainder. Condition designation for each assignment was delivered through the IDE plugin. Each time a designated project was opened in the IDE, students were reminded to enable notification suppression through the app on their phone.

Students retained full control over their phone settings and could disable the application at any time. This reduced reliance on purely voluntary self-selection of assignments while preserving realistic student behaviors.

\subsection{Measures}

\subsubsection{Programming Breaks}
A programming break is defined as a period of at least five minutes without recorded keystrokes. This is based on prior work that established that, for their university, CS1 students had a 50\% probability of having disengaged with a CS1 programming assignment after about 3.5 minutes of keystroke inactivity and the probability continues to increase thereafter ~\cite{edwards2022practical}. Break rate is computed as the number of breaks per 1,000 keystrokes, providing a measure that normalizes for active programming effort rather than just elapsed time.

Which conditions were assigned to which participants was decided pseudo-randomly. The probability of assigning the ON condition for an assignment for a student is given as
\[
p = 50 - \sum_{i=1}^{m}\frac{50}{2^i} + \sum_{j=1}^{n}\frac{50}{2^j}
\]
where $m$ is the number of assignments the student has completed in the ON condition and $n$ is the number of assignments in the OFF condition. For example, if the student has completed 2 ON assignments and 2 OFF assignments, the probability of assigning the next assignment as ON would be $50\%$. If the student has completed 3 ON assignments and 1 OFF assignment, the probability would be $12.5\%$.


\subsection{Statistical Analysis}
The primary measure we use is programming break count at the student-assignment-condition level. Because assignments varied in programming effort, break counts are modeled relative to total keystroke exposure.

Our primary analysis uses a generalized linear mixed model (GLMM) with a negative binomial distribution. Break count is the dependent variable, the log of total keystrokes is included as an offset, and notification suppression status is included as a fixed effect. Random intercepts for students and assignments account for repeated measures and assignment-level heterogeneity. Exponentiation coefficients are interpreted as incidence rate ratios (IRR).





\section{Results}

This section reports descriptive and model-based analyses examining the relationship between phone-notification suppression and programming behavior. The results are organized to first establish assignment-level exposure and descriptive patterns in break rates, followed by mixed-effects modeling and an examination of variability between students and assignments.

\subsection{Dataset and Exposure}

After processing and filtering to student-assignment-condition, the dataset comprised 201 student-assignment-condition observations across 22 students and 8 programming assignments. Each observation corresponds to a student's work on a particular assignment under a specific notification suppression status (ON or OFF), with associated counts of total keystrokes and programming breaks.


Across assignments, both ON and OFF conditions were represented, enabling within-student comparisons. Students had observations in both conditions, allowing models to estimate condition effects while accounting for individual baseline differences in breaking behavior.

Assignments differed in length and programming effort, therefore break counts were analyzed relative to total keystroke exposure. Models incorporated break rate per 1,000 keystrokes or a count-based specification with a log keystroke exposure offset. This ensures that differences in break behavior are interpreted relative to programming activity rather than raw time or assignment duration.


\begin{figure}
    \centering
    \subfloat[]{
      \includegraphics[width=.5\linewidth]{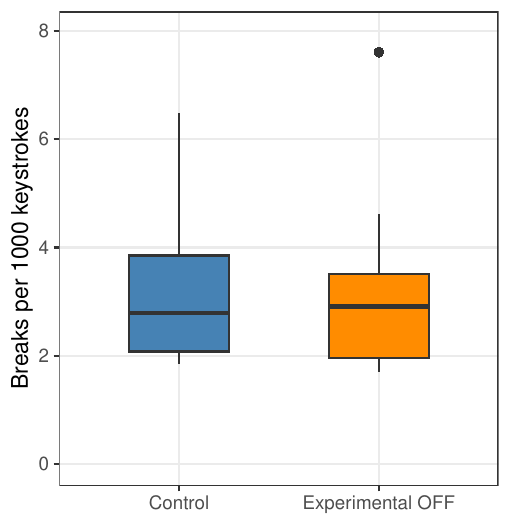}
      \label{fig:boxplotA}
    }
    \subfloat[]{
      \includegraphics[width=.5\linewidth]{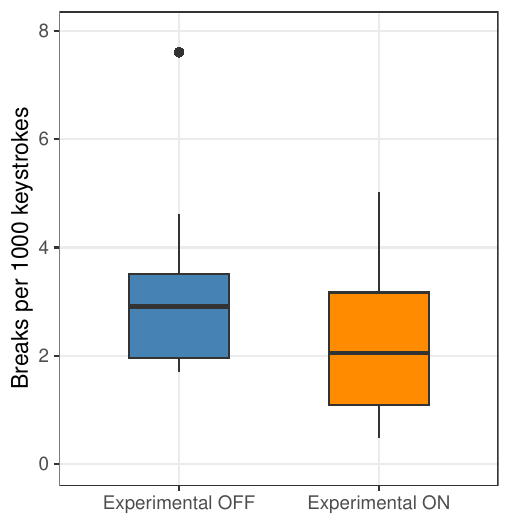}
      \label{fig:boxplotB}
    }
    \caption{\textbf{}\protect\subref{fig:boxplotA} Break rates (Breaks per 1,000 keystrokes) for non-participating control students compared to participating students during assignments without notification suppression. The similar distributions indicate that students who opted into the study had a baseline break behavior comparable to the broader class. \protect\subref{fig:boxplotB} Within participating students, break rates are generally lower when notification suppression (ON) is enabled compared to OFF assignments, suggesting an association between supression and reduced interruption frequency.}
    \label{fig:boxplots}
\end{figure}


\subsection{Descriptive Patterns}
Prior to any formal modeling, descriptive analyses were conducted to examine break-rate patterns across conditions. Break rate was defined as the number of breaks per 1,000 keystrokes, where a break is a pause of at least five minutes without recorded keystrokes.

Figure \ref{fig:boxplotA} compares break rates for students who participated in the ON condition during non-ON assignments with those of passive students who did not participate in the suppression data collection. Break rate distributions for these two groups were broadly comparable, suggesting that baseline programming behavior while not using suppression was similar across participants and the control group. In other words, students who self-selected into the study appeared to have similar break rates to those who did not actively participate in the study.

In Figure \ref{fig:boxplotB} we see that participating students'  break rates were generally lower when they were using notification suppression.

Break rates were also examined at the assignment level. Considerable variation in break rates was observed across assignments, reflecting differences in task structure, difficulty, and student engagement. Both suppression and non-suppression observations were present across assignments, indicating that the observed patterns were not solely driven by a small number of atypical tasks.

Figure \ref{fig:boxplots} shows the break‑rate distributions. The control and experimental groups are comparable in the DND‑OFF condition, while the experimental group shifts lower when DND is ON.

\subsection{Model-Based Analysis}

To account for clustering within students and assignments and to model break counts relative to programming exposure, we fit a generalized linear mixed model with a negative binomial distribution and a log keystroke exposure offset.

The fixed-effect estimate for the ON condition was $-0.210 (SE = 0.078, p=0.0068)$. With the coefficients being on the log-rate scale, exponentiation yields an incidence rate ratio (IRR = 0.81), corresponding to an estimated 18.9\% reduction in break rated under notification suppression.

\subsection{Variation Across Assignments}
In addition to the overall population-level reduction in break rates under notification suppression, meaningful variation was observed across programming assignments. The negative binomial mixed-effects model estimated non-trivial assignment-level random intercept variance, indicating that baseline break rates differed substantially by assignment. These differences are consistent with variation in assignment structure, complexity, and cognitive load.

\begin{figure}
    \centering
    \includegraphics[width=1.0\columnwidth]{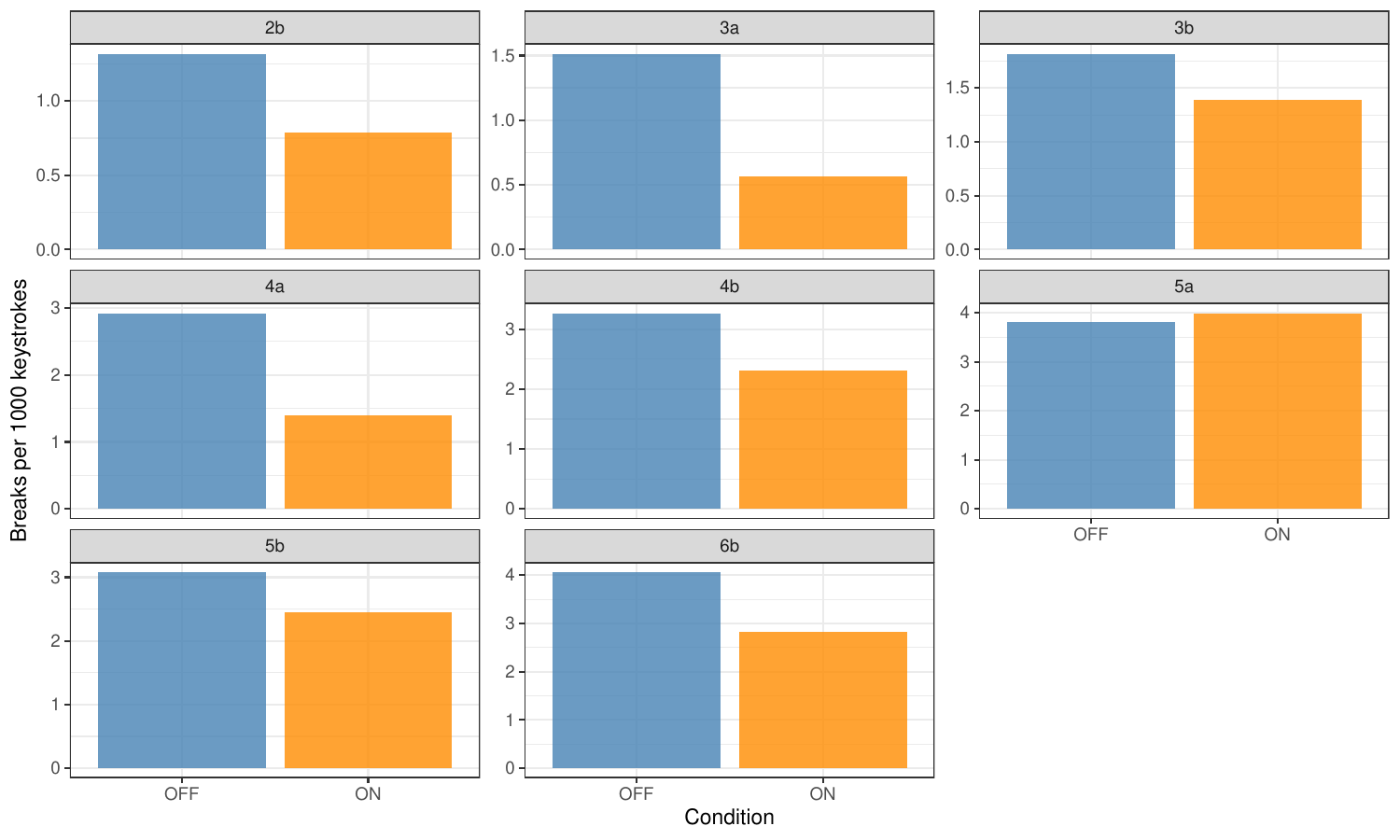}
    \caption{Break rates (breaks per 1,000 keystrokes) are shown separately for OFF and ON conditions across eight programming assignments. While several assignments show noticeable reductions in break rates under suppression, others exhibit minimal change, indication that the effect of notification suppression varies by task.}
    \label{fig:break_rates_per_assn}
    \Description{Still need to add description}
\end{figure}

Figure ~\ref{fig:break_rates_per_assn} shows within-assignment break rates under suppression ON and OFF conditions. The magnitude of the suppression effect varies across assignments, with some assignments exhibiting pronounced reductions and others showing little to no difference. This pattern supports RQ2, suggesting that the effect of notification suppression is not uniform across tasks.

\subsection{Variation Across Students}
\begin{figure}
    \centering
    \includegraphics[width=1.0\columnwidth]{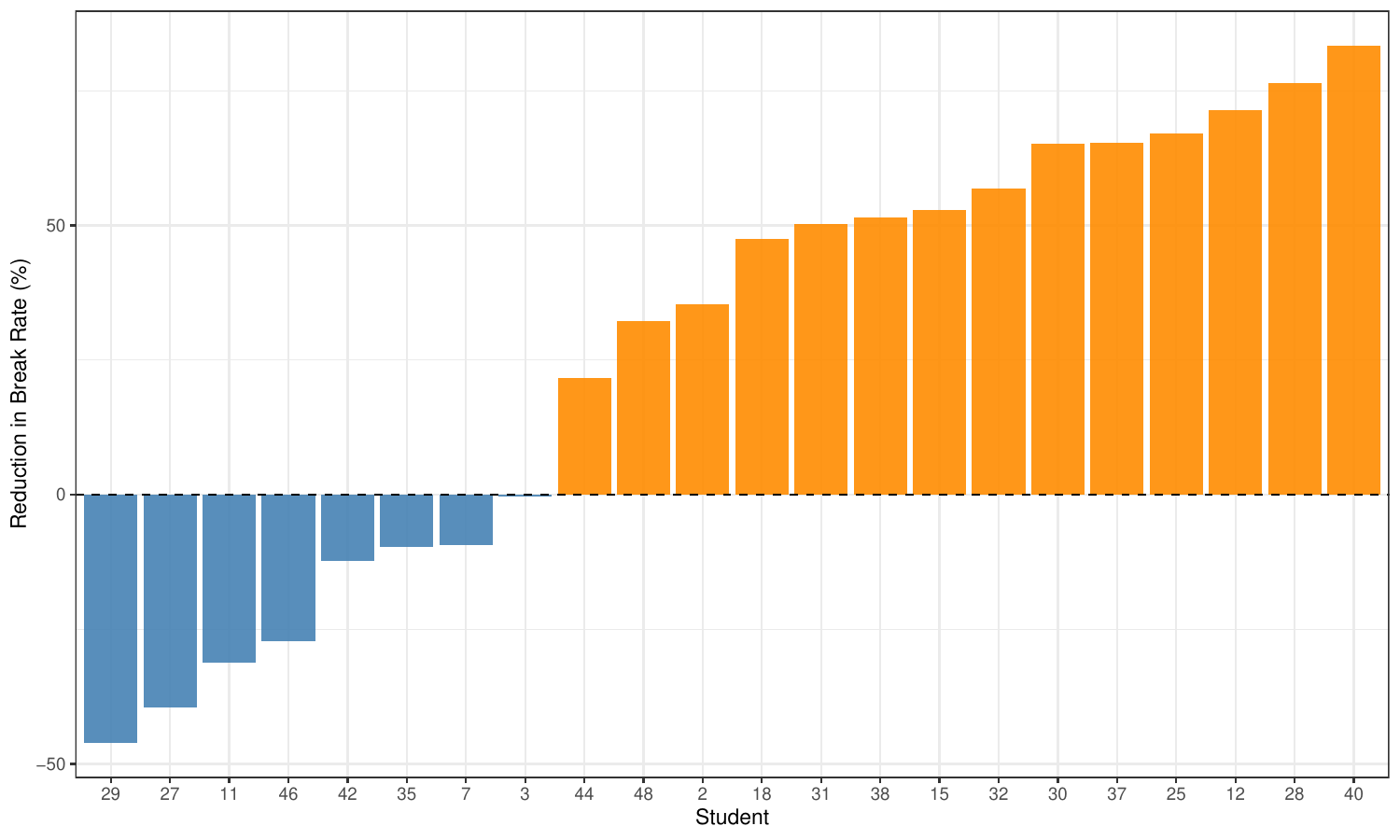}
    \caption{Each bar represents a student's percent change in break rate when notification suppression is enabled relative to when it is disabled. Most students exhibit reductions in break rates (positive values), though a minority show little change or modest increase. This illustrates individual heterogeneity despite the overall population reduction.}
    \label{fig:within_student_dnd_effect}
    \Description{Still need to add description}
\end{figure}

While the mixed-effects models estimate an average reduction in break rate under suppression, the effect is not uniform across individuals. Figure \ref{fig:within_student_dnd_effect} displays the within-student change in break rate. A majority of students exhibit reductions under suppression, though a smaller subset show negligible change or modest increases.

\begin{figure}
    \centering
    \includegraphics[width=1.0\columnwidth]{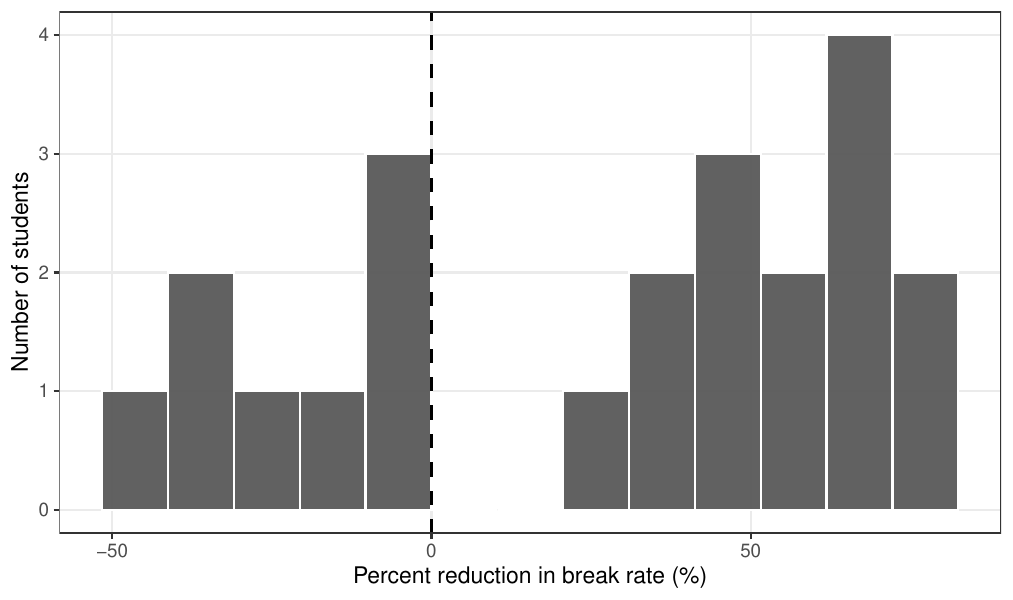}
    \caption{Histogram of percent reduction in break rate for each of the participating subjects when notification suppression relative to when it was disabled. Positive values indicate fewer breaks under suppression, while negative values indicate increases. Most students exhibit reductions, though a smaller subset show little change or modest increases.}
\label{fig:between_student_kde_pct_reduction}
    \Description{Still need to add description}
\end{figure}

Figure \ref{fig:between_student_kde_pct_reduction} visualizes the distribution of within-student percent changes. The distribution appears bimodal, with a larger mode centered around substantial reductions (approximately 50--60\%) and a smaller mode reflecting modest increases. This suggests the presence of two distinct response profiles: a majority group the benefits meaningfully from suppression and a minority group for whom suppression may not reduce interruption behavior.


\begin{figure*}
    \centering
    \includegraphics[width=0.75\textwidth]{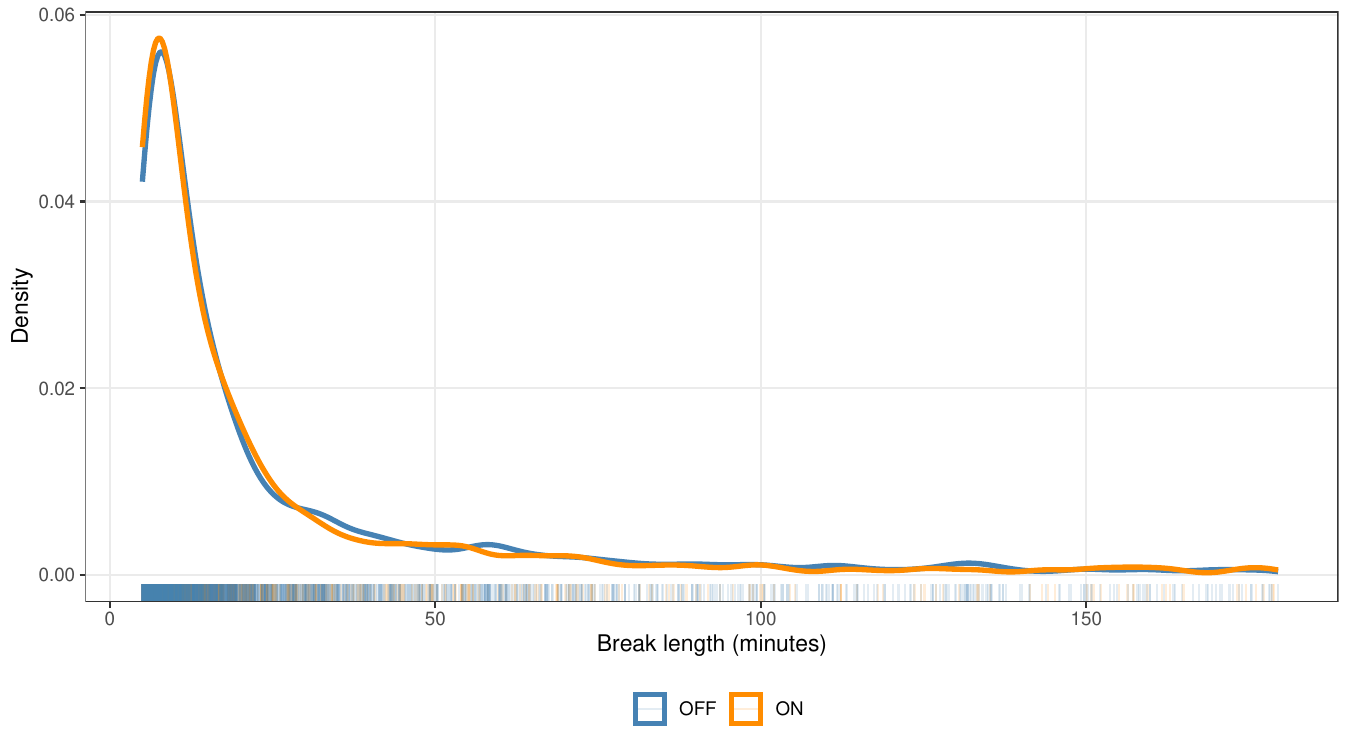}
    \caption{Kernel density estimates of break duration (breaks $\geq5$ minutes, truncated at 180 minutes) under suppression OFF and ON. The nearly identical distributions indicates that notification suppression does not meaningfully change the duration of breaks once they occur.}
\label{fig:break_length_kde}
    \Description{Still need to add description}
\end{figure*}

Although break frequency differs by condition, break duration does not. Figure \ref{fig:break_length_kde} shows nearly identical density curves for break lengths under ON and OFF conditions. This indicates that suppression primarily reduces the occurrence of breaks rather than the duration of disengagement once a break has begun.

\section{Discussion}
\subsection{RQ1: \rqone{}}

\subsection{RQ2: \rqtwo{}}

\subsection{Summary of Findings}
This study examined whether enabling phone notification suppression during programming assignments was associated with changes in programming break behavior. Using keystroke-level logs to aggregate student-assignment conditions, we modeled break counts relative to total programming activity. Across model specifications, notification suppression was associated with a reduction in programming breaks.

A negative binomial mixed-effects model indicated that notification suppression was associated with a $\approx 19\%$ reduction in breaks after accounting for student and assignment variability. Linear mixed-effects models yielded consistent results. These findings suggest that notification suppression is associated with measurable differences in programming engagement behavior.

\subsection{Variability in Response to Notification Suppression}
Although the population-level effect was negative, individual responses varied. 14 of 22 students exhibited lower break rates when using notification suppression, while 8 showed no change or increases.

The heterogeneity suggests that notification suppression does not uniformly benefit all learners. For some students, programming breaks may be driven less by phone notifications and more by other environmental factors.

Understanding which students benefit most from notification supression remains an important direction for future work.

\subsection{Implications for Programming Education}
Programming tasks require sustained cognitive engagement. Frequent interruptions, whether externally triggered or internally motivated, may fragment attention and increase context-switching. Our findings suggest that reducing external disruptions through phone-notification suppression is associated with fewer prolonged pauses in programming activity.

From an instructional perspective, this raises the possibility that lightweight behavioral interventions, such as encouraging students to enable Do Not Disturb (DND) during programming sessions, could support focus without requiring any redesign of curriculum. Because DND is built into modern mobile operating systems, it represents a scalable and low-cost intervention.



\section{Threats to validity}

\subsection{Internal Validity}
Although a within-students design was used to compare programming behavior under different notification suppression conditions, students retained full control over whether and when to enable the application. As a result, application usage may be correlated with unobserved factors such as motivation, task planning, or availability of time, which could also influence break behavior.

\section{Conclusions}
This study examined the correlations between phone notification suppression and programming break behavior in an undergraduate computer science course. Phone notifications, by design, interrupt activity to draw a user's attention to their phone. At best, phone notifications break concentration and flow, and at worst they foster a dependency on the phone. Quantitative studies of the effects of phone notifications on work productivity are challenging because such effects are generally either measured in an artificial setting or they rely on self-report. Our study measures these effects in an ecologically valid manner through unobtrusive keystroke logging while students work in their natural environment.

Our findings serve two purposes. The first is to strengthen the claims made by~\citet{hart2024phone}. These claims are that roughly $3/4$ of students take fewer breaks from programming when phone notifications are suppressed. The earlier work suffered from a small sample size; with our much-larger sample size and improved methodology, we found strong evidence that $2/3$ of students took fewer breaks with notification suppression turned on. Of particular interest, we found a bimodal distribution of effect -- while $2/3$ of students experienced strong positive effects, $1/3$ experienced mild negative effects, with very few experiencing little effect.

This result has important implications. At the most superficial, practical level, our work indicates that instructors should encourage students to turn on their phone's do-not-disturb functions while working on computer programming assignments. For most students, this should increase efficiency and could allow them longer stretches of concentration and possibly deeper flow states. However, our study also raises deep and critical questions about students who exhibit negative effects. Whether these negative effects are the result of separation anxiety, fear of missing out, and/or point to phone addiction is an important area of further study.

Our results were obtained in a CS1 course, but they are likely transferable to other contexts, both educational and professional. Our method of measuring engagement and efficiency with keystroke logging is not suitable for all contexts, but could be extended beyond coding assignments, for example, to collect data during writing natural language and technical writing.

\newpage
\bibliographystyle{abbrv}
\bibliography{paper}
\end{document}